\documentclass[a4paper,11pt]{article}
\usepackage{pos}

\title{The Role of Acoustic Instability in Cosmic-Ray Self-Confinement}
\author*[a,b]{Antonio Capanema}
\author[a,b]{Pasquale Blasi}
\author[a,b]{Emanuele Sobacchi}

\affiliation[a]{Gran Sasso Science Institute (GSSI),\\
  Viale Francesco Crispi 7, L'Aquila, Italy}

\affiliation[b]{INFN -- Laboratori Nazionali del Gran Sasso (LNGS)\\
Via Giovanni Acitelli 22, L'Aquila, Italy}

\emailAdd{antonio.capanema@gssi.it}

\abstract{
Over the past decades, there has been growing observational and theoretical evidence that cosmic-ray-induced instabilities play an important role in both acceleration and transport of cosmic rays (CRs). For instance, the efficient acceleration of charged particles at supernova remnant shocks requires rapidly growing instabilities, so much so that none of the proposed processes seem sufficient to warrant acceleration to PeV energies.

In this work, we investigate whether an acoustic instability triggered by the presence of a CR pressure gradient can lead to significant self-confinement of charged particles in the vicinity of shocks. We validate the expected growth rates and obtain the scale and energy of magnetic field perturbations induced by such system using magnetohydrodynamical (MHD) simulations. Our results suggest a strong suppression of the diffusion coefficient for particles with Larmor radius around a thousandth of the precursor scale length.
}

\ConferenceLogo{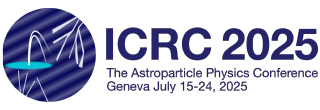}

\FullConference{39th International Cosmic Ray Conference (ICRC2025)\\
 15–24 July 2025\\
Geneva, Switzerland\\}
\begin{document}
\maketitle

%%%%%%%%%%%%%%%%%%%%%%%%%%%%%%%%%%%%%%%%%%%%%%%%%%%%%%%%%%%%%%
\section{Introduction}
\label{sec:intro}

Over the past decades, it has become well established that an accurate description of cosmic-ray (CR) transport requires abandoning the usual test-particle approach, thereby accounting for their own dynamical feedback onto the system. These effects are expected to be crucial at the vicinity of shock waves, which are key sites of CR acceleration across a wide range of astrophysical settings. This is because explaining the highest energy Galactic and extragalactic CRs requires more efficient shocks and stronger magnetic fields than standard diffusive shock acceleration can predict. In light of this issue, CR-induced resonant \cite{Skilling:1975kf} and non-resonant/Bell \cite{Bell:2004hhd} streaming instabilities gained popularity due to their potential to rapidly amplify magnetic fields, suppressing the diffusion coefficient and enhancing particle acceleration at shock fronts \cite{Bell:1978dv}. Alas, reaching the CR knee's $\sim$~PeV energies at supernova remnant shocks is likely too tall of an order for such mechanisms \cite{Cristofari:2021hbc}.

In this work, we explore a different mechanism that can also inhibit CR diffusion around astrophysical shocks: the so-called ``acoustic'' or ``Drury instability'' \cite{Drury:1984do,Drury:1986vd}. This instability is triggered by the pressure gradient produced by CRs diffusing at shock precursors, which interact with inhomogeneities present in the upstream plasma, leading to their exponential growth while approaching the shock surface. In \cite{Beresnyak:2009pi} and \cite{Drury:2012xb}, important steps were taken toward understanding its magnetic-field amplification and turbulence generation, capabilities. These works showed that magnetic field perturbations can get amplified to nonlinear levels (\textit{i.e.} comparable to or larger than the background field), strongly suppressing diffusion at supernova remnant shock precursors, potentially reaching $\sim 10$~PeV CR energies \cite{Beresnyak:2009pi}.

After deriving the growth rate of the Drury instability in Section \ref{sec:instability}, we perform magnetohydrodynamical (MHD) simulations of a CR-modified shock precursor region in order to confirm our expectations in Section \ref{sec:sims}. Section \ref{sec:nonlinear} is then dedicated to studying the nonlinear stage of the system's evolution, with a particular focus on magnetic field amplification. We finalize by briefly comparing the Drury and Bell growth rates and saturation scales, discussing the roles of each one for CR acceleration.

%%%%%%%%%%%%%%%%%%%%%%%%%%%%%%%%%%%%%%%%%%%%%%%%%%%%%%%%%%%%%%
\section{Drury Instability}
\label{sec:instability}

To derive the instability, we consider a plasma with density $\rho$, velocity $\mathbf{u}$, pressure $P$, and adiabatic index $\gamma$ evolving according to the equations of hydrodynamics (HD), coupled to an external force induced by the CR pressure gradient, $\nabla P_{\rm CR}$:
\begin{equation}\label{eq:HDeqs}
    \frac{\partial \rho}{\partial t} + \nabla\cdot (\rho \mathbf{u}) = 0~, \quad
    \frac{\partial \mathbf{u}}{\partial t} + (\mathbf{u}\cdot\nabla)\mathbf{u} = -\frac{\nabla P}{\rho} - \frac{\nabla P_{\rm CR}}{\rho}~, \quad
    \frac{\partial}{\partial t}\left(\frac{P}{\rho^\gamma}\right) + (\mathbf{u}\cdot \nabla)\left(\frac{P}{\rho^\gamma}\right) = 0~.
\end{equation}
We begin by perturbing a uniform and static background solution, $\rho=\rho_0 + \delta\rho$, $\mathbf{u}=\delta\mathbf{u}$, and $P=P_0 + \delta P$, obtaining the linearized equations 
\begin{equation}\label{eq:HD-perturb}
    \frac{\partial \delta\rho}{\partial t} + \rho_0 \nabla\cdot \delta \mathbf{u} = 0~,\quad
    \frac{\partial \delta \mathbf{u}}{\partial t} = -\frac{\nabla \delta P}{\rho_0} + \frac{\nabla P_{\rm CR}}{\rho_0^2}\delta\rho~,\quad
    \frac{\partial}{\partial t}\left(\frac{\delta P}{P_0} - \gamma\frac{\delta \rho}{\rho_0}\right) = 0~.
\end{equation}
By assuming the perturbations are $\propto\exp[i(\mathbf{k}\cdot\mathbf{x}-\omega t)]$, one finds the following dispersion relation:
\begin{equation}\label{eq:dispersion}
    \omega^2 = c_s^2 k^2 + \frac{i\mathbf{k}\cdot \nabla P_{\rm CR}}{\rho_0}~,
\end{equation}
where $c_s = \sqrt{\gamma P_0/\rho_0}$ is the sound speed of the unperturbed medium. The imaginary part of $\omega$ characterizes the (exponential) growth rate of perturbation amplitudes in the Drury instability. Although this derivation is much simpler than the rigorous WKB-like one presented in DF, it still captures its essential physics, at the cost of missing a couple of minor effects. Specifically, we overlook (\textit{i}) the damping due to CR friction \cite{Ptuskin:1981br}, which is negligible at shock precursors, and (\textit{ii}) the back-reaction of the instability onto the CRs themselves, which we leave for a future study.

There are two regimes characterizing the growth rate of the Drury instability:
\begin{equation}\label{eq:growth-rate}
    \Gamma \to\begin{cases}
        \sqrt{\frac{k|\nabla P_{\rm CR}|}{2\rho_0}}~, & k\ll \frac{|\nabla P_{\rm CR}|}{\rho_0 c_s^2}\\
        \frac{|\nabla P_{\rm CR}|}{2\rho_0 c_s}~, & k\gg \frac{|\nabla P_{\rm CR}|}{\rho_0 c_s^2}
    \end{cases}~.
\end{equation}
Notably, the maximum growth rate at large $k$ is $k$-independent, while at small $k$ it becomes $\propto\sqrt{k}$. The transition between both regimes can be written as
\begin{equation}
    {\rm Re}(\omega^2) \sim |{\rm Im}(\omega^2)| \qquad\Rightarrow\qquad c_s^2k \frac{\delta\rho}{\rho_0} \sim \frac{|\nabla P_{\rm CR}|}{\rho_0}\frac{\delta\rho}{\rho_0}~,
\end{equation}
which translates to both terms on the right-hand side of the middle equation in (\ref{eq:HD-perturb}) being comparable. In the long-wavelength (small-$k$) limit, the acceleration $\partial\delta\mathbf{u} /\partial t$ of a perturbed fluid element is dominated by the CR pressure gradient term, while in the small-wavelength (large-$k$) limit it is dominated by the fluid's own pressure gradient term. 

The introduction of magnetic fields affects the Drury instability, depending on their orientation with respect to $\nabla P_{\rm CR}$. A similar procedure of perturbing the ideal MHD equations under a uniform background ($\rho_0$, $\mathbf{u}_0=0$, $P_0$, $\mathbf{B}_0$) and perturbations $\delta\rho,\delta\mathbf{u},\delta P,\delta\mathbf{B}\propto e^{i(\mathbf{k}\cdot\mathbf{x}-\omega t)}$ yields
\begin{equation}
    \mathbf{B}=\mathbf{B}_0+\delta\mathbf{B}
\end{equation}
\begin{align}
    \omega^2 &= c_s^2 k^2 + \frac{i\mathbf{k}\cdot \nabla P_{\rm CR}}{\rho_0}~, \qquad (\mathbf{k}\parallel \mathbf{B}_0)\label{eq:kpar-dispersion} \\
    \omega^2 &= c_{ms}^2 k^2 + \frac{i\mathbf{k}\cdot \nabla P_{\rm CR}}{\rho_0}~, \quad\,\; (\mathbf{k}\perp \mathbf{B}_0)\label{eq:kperp-dispersion}
\end{align}
where $c_{ms}^2 = c_s^2 + v_A^2$ and $v_A=B_0/\sqrt{4\pi\rho_0}$ are the unperturbed magnetosonic and Alfvén speeds, respectively. The dispersion relations (\ref{eq:kpar-dispersion}) and (\ref{eq:dispersion}) are identical, meaning that a parallel background field does not affect the instability. On the other hand, a perpendicular field slows down the instability's maximum growth rate by changing all instances of $c_s$ in equation (\ref{eq:growth-rate}) into $c_{ms}$.

\begin{figure}[t!]
    \centering
    % \sidecaption
    \includegraphics[width=0.7\linewidth]{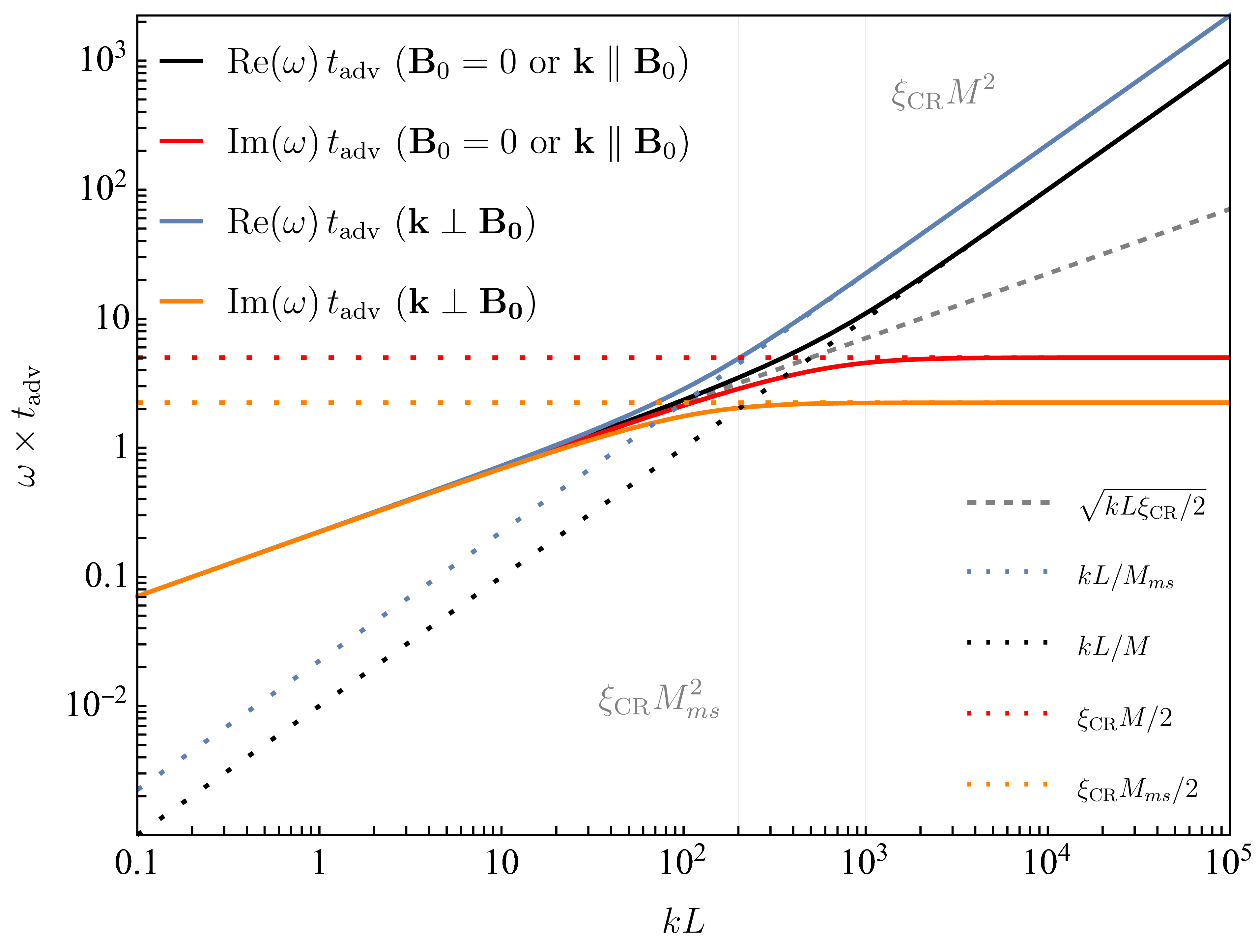}
    \caption{Real and imaginary parts of $\omega$ (multiplied by $t_{\rm adv}=L/u_0$) from relations (\ref{eq:dispersion})/(\ref{eq:kpar-dispersion}) and (\ref{eq:kperp-dispersion}), assuming $|\nabla P_{\rm CR}|=\xi_{\rm CR}\rho_0u_0^2/L$ parallel to $\mathbf{k}$. Parameter values are chosen to be $\xi_{\rm CR}=0.1$, $M=100$, and $M_{ms}=100/\sqrt{5}$. Asymptotic behaviors are represented by the gray dashed and colored dotted lines.}
    \label{fig:dispersions}
\end{figure}

Figure \ref{fig:dispersions} shows the real and complex parts of $\omega$ as functions of $k$, as predicted by equations (\ref{eq:kpar-dispersion}) and (\ref{eq:kperp-dispersion}), along with their asymptotic expressions (dashed and dotted lines) and values of $k$ where there is a transition between regimes (gray vertical lines). In analogy with the simulations from the following sections, we have parametrized the CR pressure gradient as $|\nabla P_{\rm CR}|=\xi_{\rm CR}\rho_0u_0^2/L$, representing a shock wave precursor region, with CRs producing a fraction $\xi_{\rm CR}$ of the ram pressure at the shock, and with a constant gradient over a length $L$. We have also multiplied all quantities by a reference advection time $t_{\rm adv}=L/u_0$, such that the number of growth e-folds is $\Gamma t_{\rm adv}=\sqrt{kL\xi_{\rm CR}/2}$ in the $kL\ll \xi_{\rm CR}M^2$ regime without magnetic fields, where $M=u_0/c_s$ is the shock Mach number, and saturating at a maximum value for $kL\gg \xi_{\rm CR}M^2$:
\begin{equation}\label{eq:max-e-folds}
    \text{Maximum~number~of~e-folds} = \frac{\xi_{\rm CR} M}{2}~.
\end{equation}
In the presence of a perpendicular magnetic field, the sonic Mach number $M$ should be replaced with the magnetosonic one $M_{ms}=u_0/c_{ms}$. Adopted values were $\xi_{\rm CR}=0.1$, $M=100$ for a parallel field or no field (red and black lines), and $M_{ms}=100/\sqrt{5}$ for a perpendicular field (blue and orange lines). In most optimistic scenario with such parameters, we predict that perturbations should grow in amplitude by a factor of $\exp(\xi_{\rm CR}M/2) = \exp(5) \approx 150$ in one advection time.

%%%%%%%%%%%%%%%%%%%%%%%%%%%%%%%%%%%%%%%%%%%%%%%%%%%%%%%%%%%%%%
\section{Simulations: Linear Regime}
\label{sec:sims}

In order to test the validity of the expressions obtained in the previous sections, we perform MHD simulations using the PLUTO\footnote{For the numerical details of our simulations, we refer the reader to the original article, in preparation.} code \citep{Mignone:2007iw}. Our 2D simulation box has dimensions $(x,y) \in [0,L]\times [0,L/8]$, split into of a uniform discrete grid with $N_x \times N_y$ cells, small enough to avoid numerical dissipation. This region represents the precursor of a shock wave in its rest frame, with the shock lying just beyond the right boundary at $x=L$. At $x=0$, we have ``upstream infinity'', where the external medium's plasma with MHD fluid variables $\rho_0$, $P_0$, $\mathbf{B}_0$, and adiabatic index $\gamma=5/3$ is entering the simulation box at the shock speed $\mathbf{u}_0=u_0\,\hat{\mathbf{x}}$.

We introduce the CR pressure gradient 
\begin{equation}
    P_{\rm CR}(x) = \xi_{\rm CR}\,\rho_0u_0^2\,\frac{x}{L} \quad\Rightarrow\quad \nabla P_{\rm CR}(x) = \xi_{\rm CR}\,\frac{\rho_0u_0^2}{L}\,\hat{\mathbf{x}}~.
\end{equation}
as a body force $\mathbf{g}=-\nabla P_{\rm CR}/\rho$ under the ``\texttt{VECTOR}'' prescription in PLUTO. Its presence induces non-uniform density, velocity, pressure, and magnetic field profiles along $x$ at steady state. Their equations are obtained by solving the MHD equations with the added $\nabla P_{\rm CR}$ term, and then solved numerically to be used as the initial conditions for all simulations. Boundary conditions are set to periodic at $y=0$ and $y=L/8$, and to outflow at $x=L$. At $x=0$, we set
\begin{equation}\label{eq:BC}
    \rho(x=0,y,t) = \rho_0+\delta\rho_0\sin(k_x u_0 t + k_y y + \phi)~,
\end{equation}
with $\phi$ being a random phase in the interval $[0,2\pi]$, along with $\mathbf{u}=u_0\, \hat{\mathbf{x}}$, $P=P_0$, and either $\mathbf{B}=B_{x,0}\,\hat{\mathbf{x}}$ or $\mathbf{B} = B_{y,0}\,\hat{\mathbf{y}}$. This corresponds to a continuous inflow of density fluctuations into the system.

Code units in PLUTO are dimensionless and chosen such that $L=1$, $\rho_0=1$, and the sound speed at upstream infinity is $c_{s,0}=1$. As a result, in dimensionless code units, the pressure at upstream infinity is fixed to $P_0= 0.6$, while $u_0$, $\delta\rho_0$, $B_0$, and $\lambda_{x,y}$ in code units become numerically equivalent to the dimensionless quantities $M=u_0/c_{s,0}$, $\delta\rho_0/\rho_0$, $v_{A,0}/c_{s,0}$, and $L\lambda_{x,y}^{-1}$ respectively, which are more generic and useful for applications to real physical systems.

\begin{figure*}[t!]
    \centering
    \includegraphics[width=\linewidth]{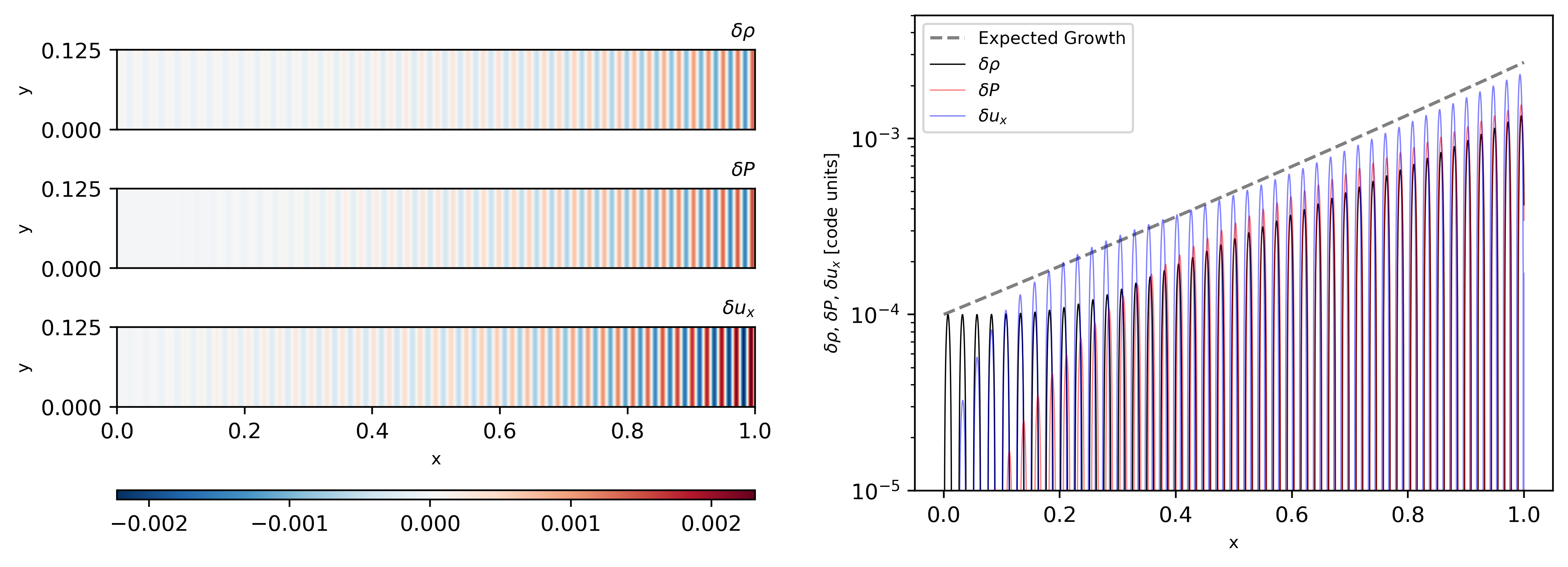}
    \caption{\textit{Left:} Steady-state perturbation profiles for $\xi_{\rm CR}=0.1$, $M=100$, $\delta\rho_0/\rho_0=10^{-4}$, $B_0=0$, $k_xL=80\pi$, and $k_y=0$. \textit{Right:} Constant-$y$ slice of profiles on the left, comparing the simulation growth of perturbations with the expected one in equation (\ref{eq:growth-rate}).}
    \label{fig:kx}
\end{figure*}

Using a grid with $N_x=4000$ and $N_y=4$, we show in the left panel of Figure \ref{fig:kx} the steady-state 2D perturbation profiles of a simulation with $\xi_{\rm CR}=0.1$, $M=100$, $\delta\rho_0/\rho_0=10^{-4}$, $B_0=0$, $k_xL=80\pi$, and $k_y=0$. They were obtained after running the simulation for longer than a few advection times through the box, and then subtracting from the final MHD fluid variables the background profiles induced by $\nabla P_{\rm CR}$. To compare the observed and expected growth rates, we plot on the right panel the values of $\delta\rho$, $\delta P$, and $\delta u_x$ at a constant-$y$ slice. We can clearly see that simulation agrees with equation (\ref{eq:growth-rate}) for this choice of parameters. We have also testes other combinations of parameters, both with and without magnetic fields; our analytical results are confirmed in all cases with $k_x$ modes.

Meanwhile, we do not expect to see any growth of $k_y$ modes, as predicted by the dispersion relations in Section \ref{sec:instability}. However, perpendicular perturbations are also affected by the CR pressure gradient, since the acceleration $-\nabla P_{\rm CR}/\rho$ of a fluid element is different for overdensities and underdensities, independently of their orientation. Namely, regions with $\rho+\delta\rho$ decelerate slightly less than those with $\rho-\delta\rho$:
\begin{equation}\label{eq:over-under}
    -\frac{\nabla P_{\rm CR}}{\rho_0 + \delta\rho_0\sin(\mathbf{k}\cdot\mathbf{x}+\phi)} \approx -\frac{\nabla P_{\rm CR}}{\rho_0} + \frac{\nabla P_{\rm CR}}{\rho_0}\frac{\delta\rho_0}{\rho_0}\sin(\mathbf{k}\cdot\mathbf{x}+\phi)~.
\end{equation}
When magnetic fields are present, magnetic tension forces also arise along the $x$-direction, since the field is frozen-in with the density. We are able to confirm the presence of both of these effects quantitatively in simulations. In the more general case of ISM inhomogeneities with both $k_x$ and $k_y$ modes, these effects are coupled with the exponential growth of MHD fluid variables due to the Drury instability.

%%%%%%%%%%%%%%%%%%%%%%%%%%%%%%%%%%%%%%%%%%%%%%%%%%%%%%%%%%%%%%
\section{Simulations: Nonlinear Regime}
\label{sec:nonlinear}

Next, we shift our focus to the nonlinear regime, characterized by large perturbations. We employ the same simulation framework as described in the previous section, with the goal to investigate different effects emerging at shock wave precursors when $\delta\rho,\delta P,\delta u, \delta B \gtrsim \rho_0,P_0,u_0,B_0$, as well as to assess the saturation scale of the CR-pressure-induced turbulence arising from the combination of these effects. In particular, previous works have posited that the process responsible for magnetic field amplification this regime is the so-called small-scale dynamo \cite{Beresnyak:2009pi}.

\begin{figure*}[t!]
    \centering
    \includegraphics[width=\linewidth]{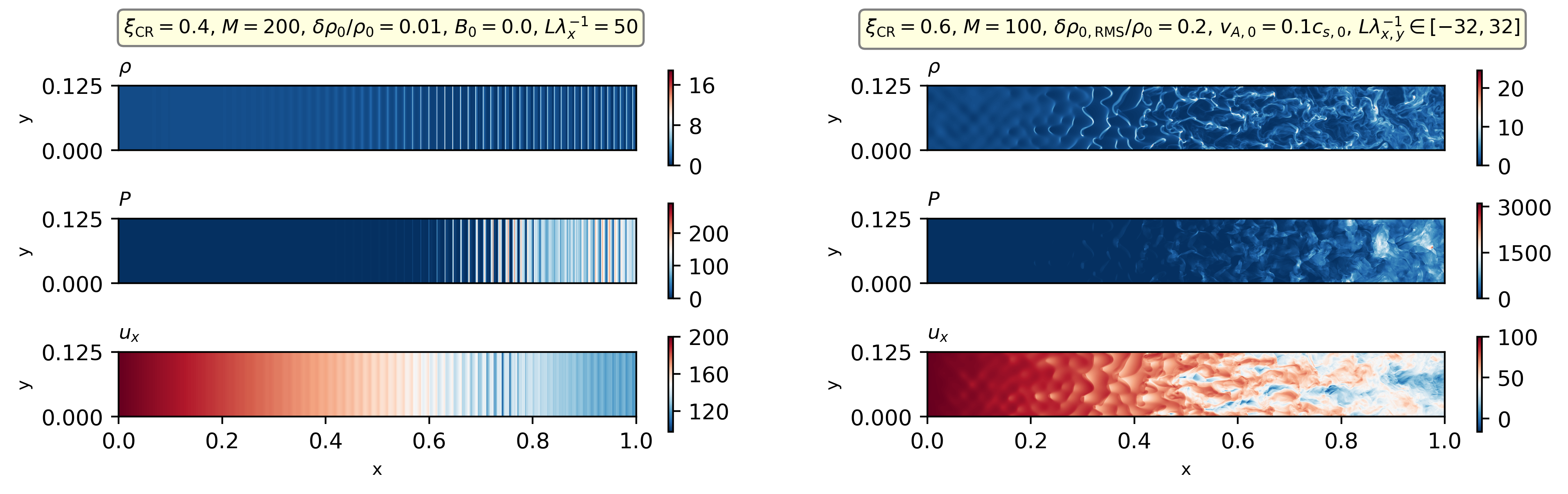}
    \caption{Density, pressure and velocity steady-state profiles in simulations where perturbations become larger than their background values. \textit{Left:} Injection of $k_x$ mode with $\delta\rho_0< \rho_0$, growing under the Drury instability until the nonlinear stage. \textit{Right:} Several modes are injected in both directions, generating turbulence and short-lived stalled regions.}
    \label{fig:nonlinear}
\end{figure*}

On the left panel of Figure \ref{fig:nonlinear}, we illustrate a situation in which a small ($\delta\rho_0/\rho_0=0.01$) $k_x$ mode enters the shock precursor and grows exponentially according to the Drury instability, becoming $\gtrsim \rho_0$ before reaching halfway through the simulation box. At this point, density contrasts are so large that $\nabla P_{\rm CR}/\rho$ makes overdensities start traveling significantly faster than underdensities, as predicted by equation (\ref{eq:over-under}). The resulting steepening of oscillations culminates in the formation of overdense walls that launch high-pressure shock waves, similar to a supersonic piston. Once these small-scale shocks reach the next overdense ``wall'', there is interference of shocked regions and formation of weak subshocks, whose influence on CRs and their acceleration was studied in \cite{Kang:1992vw}. Perhaps the most important takeaway from this panel is that the Drury instability alone is capable of bringing upstream perturbations from the linear into the strongly nonlinear regime within one advection time.

On the right, we inject many perturbation modes, similar to \cite{Drury:2012xb}:
\begin{equation}\label{eq:DDdensity}
    \rho(x=0,y,t) = \rho_0 + \alpha\sum_{\substack{a,b=0 \\ (a,b)\neq(0,0)}}^{32} A_{ab}\sin\left[\frac{2\pi}{L} (au_0t+by) + \phi_{ab}\right]~,
\end{equation}
where $A_{ab}$ and $\phi_{ab}$ are random amplitudes and phases of each perturbation drawn uniformly from $[0,1]$ and $[0,2\pi]$, respectively, and $\alpha$ is a constant chosen in order to fix $\delta\rho_{0,\rm RMS}$. In this simulation, we notice strong turbulence forming by the interference of different modes once they become nonlinear. We also observe the formation of short-lived regions with negative velocity, caused by the strong deceleration of underdense regions as predicted by equation (\ref{eq:over-under}).

\begin{figure}[t!]
    \centering
    \begin{minipage}{0.49\textwidth}
        \centering
        \includegraphics[width=\linewidth]{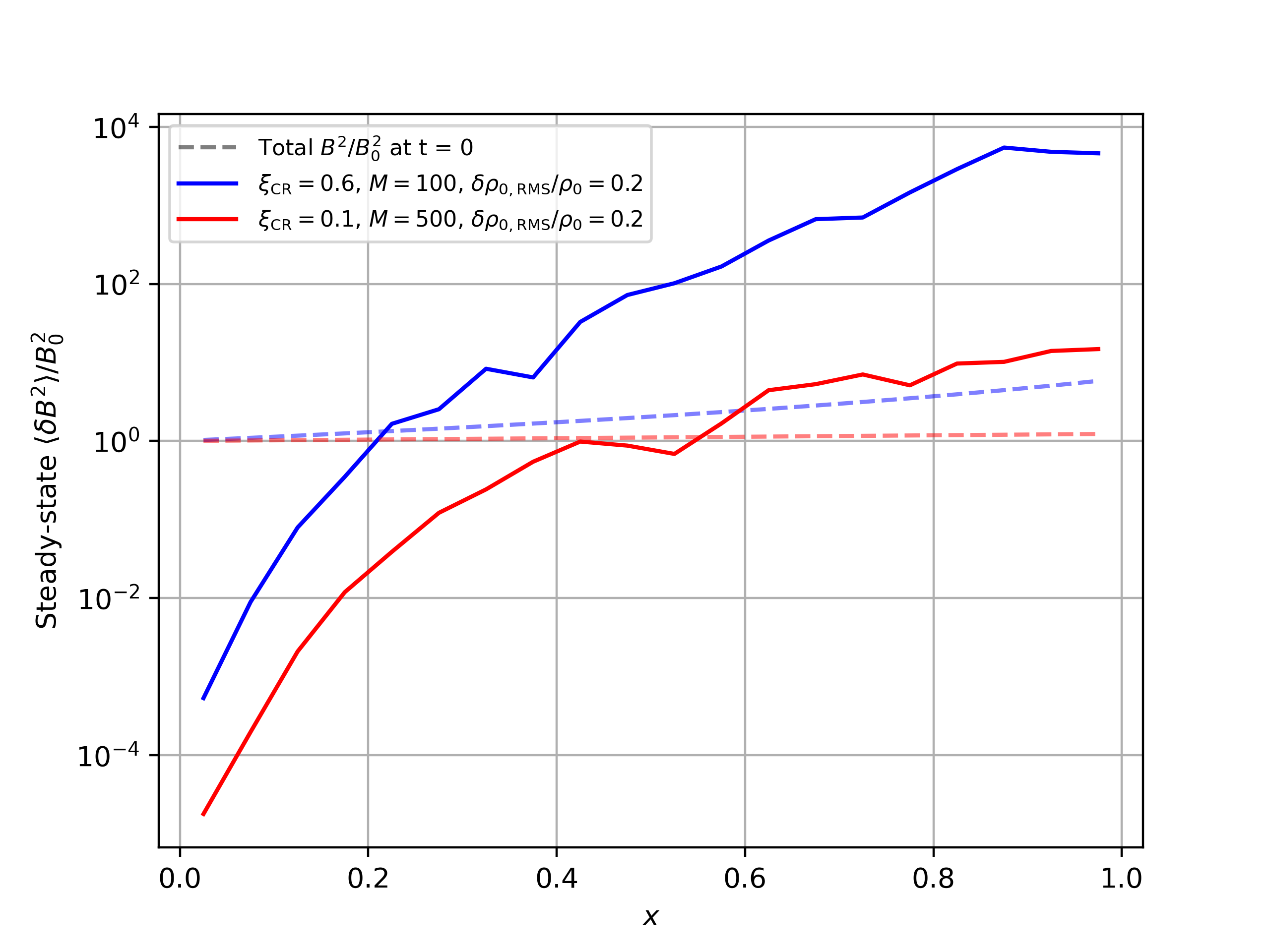}
        \caption{Amplification of the average magnetic energy in perturbations at steady state. For comparison, dashed lines represent the total magnetic energy at $t=0$, which is entirely due to mean fields that remain mostly unchanged as the simulation progresses.}
        \label{fig:energy}
    \end{minipage}
    \hfill
    \begin{minipage}{0.49\textwidth}
        \centering
        \includegraphics[width=\linewidth]{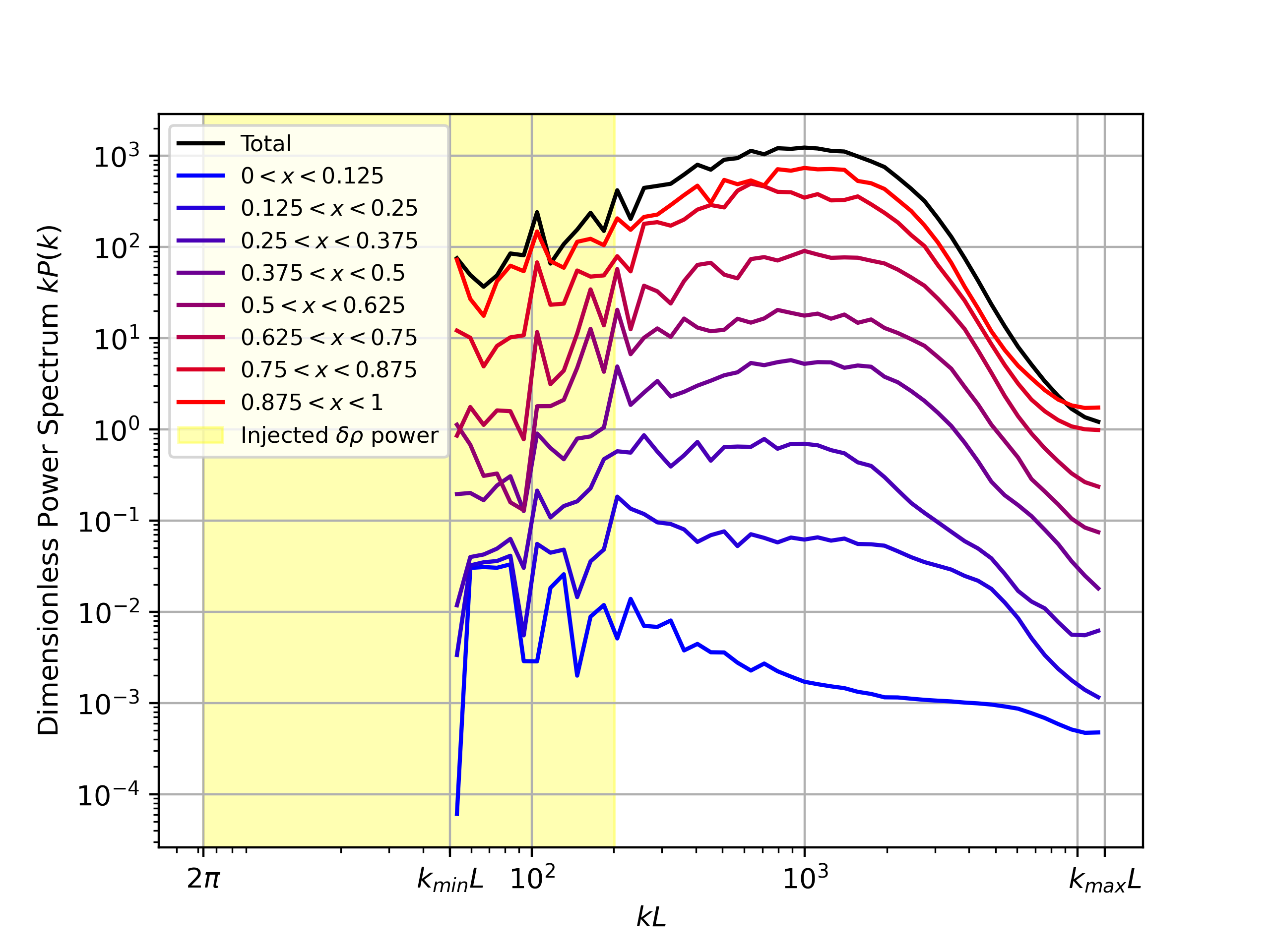}
        \caption{Dimensionless power spectrum of magnetic field perturbations at steady-state, for the simulation with $\xi_{\rm CR}=0.6$ and $M=100$. We show the total power in the whole box (black line), as well as the power in each eighth of the box (colored lines).}
        \label{fig:powerspec}
    \end{minipage}
\end{figure}

Let us briefly comment on what happens to the magnetic field in the last simulation. The mean magnetic field $\langle\mathbf{B}\rangle$ remains roughly constant ($\langle\cdot\rangle$ denotes a spatial average), both as we evolve the system in time and as we move from $x=0$ to $x=L$ at any given time. However, the total magnetic energy in the box increases by $\sim 3$ orders of magnitude as the simulation progresses towards steady state, implying that this energy is going almost exclusively into the induced field perturbations $\delta \mathbf
B=\mathbf{B}-\langle\mathbf{B}\rangle$. Figure \ref{fig:energy} shows the evolution with $x$ of the (dimensionless) average energy density in perturbations, 
\begin{equation}
    \frac{\langle u_{\delta B}\rangle}{B_0^2/8\pi}=\frac{\langle\delta B^2\rangle}{B_0^2}
\end{equation}
for that simulation, as well as for another one using the more conservative set of parameters $\xi_{\rm CR}=0.1$ and $M=500$. In the former case, the magnetic energy in the form of perturbations at $x\sim L$ saturates at $\sim 10^3$ times the total energy present there at $t=0$. This energy is mostly concentrated in modes with $kL\sim 10^3$, as revealed by its (dimensionless) power spectrum, displayed in Figure \ref{fig:powerspec}. The power spectrum also shows how the power moves from large- to small-scale modes as turbulence develops by moving along $x$. Adopting the conservative parameters, we find that the energy in $\delta B$ at $x=L$ saturates only $\sim 1$ order of magnitude above the total one energy initially there. However we stress that our results are limited by grid resolution: if a significant magnetic energy budget is present at scales $\lesssim 50L/N_x$ ($kL\sim 10^3$ in Figure \ref{fig:powerspec}, which is likely not a coincidence), numerical damping may be suppressing most of it. Another caveat is that we are neglecting the back-reaction of these effects into the CR pressure gradient, which requires more involved MHD + particle-in-cell simulations. Still, our results are quite encouraging for CR acceleration.

%On the other hand, the energy density present in the turbulence can be estimated as $u_T \approx \delta\rho \delta u^2/2$

Finally, we should discuss the role of the Bell instability within this picture. Bell is excited when the upstream-directed flux of escaping CRs produces a nonzero net current $J_{\rm CR}$ moving along with the shock, provided that $J_{\rm CR}E_{\rm CR}>qB_0^2/4\pi$, a condition we expect to be valid much further away than the precursor length $L$. The result is the amplification -- and potential saturation -- of magnetic field perturbations via Bell, which has a maximum growth rate of $\Gamma_{\rm Bell}(k_{max})=k_{max}v_A$, where $k_{max}=4\pi J_{\rm CR}/cB_0$, before the onset of Drury instability at the precursor. These large perturbations will get be further amplified via Drury instability/small-scale dynamo before reaching the shock.

\bibliographystyle{JHEP}
\bibliography{refs}

\end{document}